%% file: WP_18_Shock.tex
\documentclass[11pt,a4paper]{article}

\input{ASTROHWhitePaperStyle.tex}
\input{WP_AuthorList.tex}

\usepackage{xspace}

\begin{document}

\input{myheader}

\newcommand{\eflux}{{\rm erg~cm^{-2}\, s^{-1}}}
\newcommand{\lat}{\emph{Fermi}-LAT\xspace}
\newcommand{\fermi}{\emph{Fermi}\xspace}
\newcommand{\agile}{\emph{AGILE}\xspace}
\newcommand{\chandra}{\emph{Chandra}\xspace}
\newcommand{\xmm}{XMM-\emph{Newton}\xspace}
\newcommand{\suzaku}{\emph{Suzaku}\xspace}
\newcommand{\asca}{\emph{ASCA}\xspace}
\newcommand{\integr}{\emph{INTEGRAL}\xspace}
\newcommand{\ROSAT}{\emph{ROSAT}\xspace}
\newcommand{\sax}{\textit{Beppo}SAX\xspace}
\newcommand{\swift}{\textit{Swift}\xspace}
\newcommand{\nustar}{\emph{NuSTAR}\xspace}
\newcommand{\astroh}{\emph{ASTRO-H}\xspace}
\newcommand{\hess}{{H.E.S.S.}\xspace}
\newcommand{\pizero}{$\pi^0$-decay\xspace} 
\newcommand{\simx}{{\tt simx}\xspace}
\newcommand{\xspec}{{\tt XSPEC}\xspace}
\newcommand{\comptel}{COMPTEL\xspace}

\newcommand{\rxj}{RX~J1713.7$-$3946\xspace} 
\newcommand{\psrb}{PSR~B1259$-$63\xspace} 
\newcommand{\rxp}{2RXP~J130159.6$-$635806\xspace}
\newcommand{\hessj}{HESS~J0632$+$057\xspace}
\newcommand{\lsi}{LSI~+61\deg~303\xspace}
\newcommand{\ls}{LS~5039\xspace}
\newcommand{\fgl}{1FGL J1018.6$-$5856\xspace}
\newcommand{\cygxI}{Cyg~X-1\xspace}
\newcommand{\cygxIII}{Cyg~X-3\xspace}

\def\deg {$^\circ$}
\def\gr{$\gamma$-ray}
\def\ndash{--}

\newcommand{\WhitePaperTitle}{Shock and Acceleration }
\newcommand{\WhitePaperAuthors}{
F.~Aharonian~(DIAS \& MPI-K),
Y.~Uchiyama~(Rikkyo~University),
D.~Khangulyan~(JAXA),
T.~Tanaka~(Kyoto~University),
M.~Chernyakova~(DIAS),
T.~Fukuyama~(JAXA), and
J.~Hiraga~(University~of~Tokyo)}
\MakeWhitePaperTitle

\begin{abstract}
We discuss  the prospects for a progress to be brought by ASTRO-H in the understanding of 
the physics of  particle acceleration in astrophysical environments. 
Particular emphasis will be put on  the synergy  with $\gamma$-ray astronomy, in the context of 
the rapid developments of recent years. 
Selected topics include: shock acceleration in supernova remnants (SNRs)  
and in clusters of galaxies, and the extreme particle acceleration seen in gamma-ray 
binaries. Since the  hydrodynamics  and thermal properties  of 
shocks in these objects are covered  in  other white papers, 
we focus on the aspects related to the process of  particle acceleration.  
In the case of SNRs, we emphasize the importance of SXS and HXI observations of the X-ray
emission of  young SNRs  dominated by synchrotron radiation, particularly SNR RX J1713.7$-$3946. 
We argue that  the HXI observations of  young  SNRs, 
as a byproduct of SXS observations dedicated for 
studies of the  shock dynamics and nucleosynthesis, will provide powerful constraints on 
 shock acceleration theories. 
Also, we  discuss gamma-ray binary systems, where extreme particle acceleration 
is inferred regardless of the nature (a neutron star or a black hole) 
of the compact object.  
Finally,  for galaxy clusters, we propose  searches for 
 hard X-ray emission of secondary electrons from interactions of  ultra-high energy cosmic rays accelerated at accretion shocks. This should allow us to understand the contribution of 
galaxy clusters to the  flux of  cosmic rays above $10^{18} $~eV.  
\end{abstract}

\maketitle
\clearpage

\tableofcontents
\clearpage

\section{Shock acceleration in supernova remnants }

\subsection{Introduction} \label{SNR:Intro} 

The importance of  studying the nonthermal phenomena in SNRs lies in the facts that: 
(1) SNRs are widely believed to be the primary sources of galactic cosmic rays (GCRs) up to the so-called ``knee" around 1 PeV; 
and (2) young SNRs are perfect laboratories  to study particle acceleration and magnetic field amplification in strong nonlinear shocks. Specifically, one can learn about diffusive shock acceleration  \citep[DSA; e.g.,][]{MD01}, which  has wide applications  in astrophysics. 
Synchrotron X-ray emission produced by multi-TeV  electrons 
accelerated at SNR shock fronts  are synergetic with very high energy (VHE) 
$\gamma$-rays \citep{CTA_Xray}.  It  follows, in particular,  from  
correlations  between X-ray and  TeV $\gamma$-ray spatial distributions in the
young  SNR \rxj\ \citep{HESS_1713_Nat, HESS_1713_2}. 
Aparently, many   issues  relevant to  the X-ray domain are tightly coupled 
with  $\gamma$-ray studies of SNRs.
In what follows, we briefly describe  the key  questions  to be addressed  by \astroh. 

\subsubsection{Key question 1: Acceleration efficiency} 

The efficiency of acceleration concerns both
 the fraction of the energy transferred to nonthermal particles, and 
 the maximum energy achievable in the acceleration process.
Here we discuss the former, and the latter in the next section. 
 
While DSA theory does not firmly predict the acceleration efficiency because of poor understanding of the microphysics of injection processes, it is generally believed to be very efficient particularly at young SNR shocks. 
Recently, numerical kinetic simulations, which adopt the so-called hybrid  ``kinetic ions and fluid electrons" approach \citep{dHybrid}, have shown 
that 10--20\% of the bulk kinetic energy of parallel and quasi-parallel strong shocks can be transferred to CR ion energies \citep{CaprioliSpitkovsky13}. 
Since fully kinetic particle-in-cell simulations are computationally too demanding for current computer capabilities, the problem of the acceleration efficiency will  possibly be solved 
from first principles only in future. 

Measuring the acceleration efficiency at SNR shocks is a key issue for understanding DSA. 
(Hereafter we consider only acceleration of CR protons and electrons for simplicity.) 
Most shock acceleration theories presume that 
strong shocks place far more energy in protons than electrons, and 
the amount of shock-accelerated CR protons determines the overall efficiency of shock-acceleration. 
Recently, the characteristic spectral feature of the $\pi^0$-decay $\gamma$-rays 
resulting from 
inelastic {\it pp} collisions has been detected in middle-aged SNRs interacting with molecular clouds \citep{Pion_Fermi}. 
The $\gamma$-ray spectroscopy provides direct evidence for proton acceleration in SNRs, 
making it  possible to measure, in principle, the acceleration efficiency. 

Also, recent \emph{Fermi} observations of Cas~A, the second youngest SNR 
in the Galaxy known to date, indicate that the GeV $\gamma$-ray spectrum is of hadronic origin \citep{Fermi_CasA2,CasA_Zira13}. The GeV $\gamma$-ray emission in Tycho's SNR 
was shown to be reconcilable only with the hadronic model  \citep{LAT_Tycho,Tycho_Hadronic}. 
The total CR contents in Cas~A and Tycho are estimated as 
$W_{\rm CR} \sim 10^{50}\ \rm erg$ using the $\gamma$-ray fluxes together with the 
multiwavelength data. 
The acceleration efficiency is therefore of the order of $\sim 10\%$ in agreement with 
the SNR paradigm of the CR origin. 

Generally, X-ray line measurements are  indispensable for quantifying 
the acceleration efficiency. 
 The $\gamma$-ray luminosity scales as 
$\propto nW_{\rm CR}$, where $n$ is the mean gas density of the $\gamma$-ray-emitting 
zone. The shock downstream of a young SNR emits thermal X-rays, and their detections 
can be used to infer the gas density $n$. 
\astroh\ will be able to inform us about the acceleration efficiency in this way. 
In Section~\ref{SNR:thermal_search}, we discuss the search for thermal X-ray lines 
from synchrotron-dominated SNRs.

If shock acceleration is efficient enough, the shock dynamics can be significantly modified 
by the pressure of  accelerated CRs, resulting in a ``CR-modified shock". 
In this regime, shock heating of the downstream plasma is expected to be reduced, 
since a sizable fraction of shock energy is channeled into nonthermal particles 
\citep{Drury09}. 
\emph{Chandra} observations of 1E~0102.2$-$7219, 
a young SNR in the Small Magellanic Cloud, were used to demonstrate that 
the postshock ion temperature is indeed lower than the temperature that would be predicted 
without invoking efficient CR acceleration \citep{Hughes_1E0102}. 
Direct measurement of the postshock ion temperature with \astroh\ will offer a key  
test  for realization of  ``CR-modified shock". 

\subsubsection{Key question 2: Maximum energy achievable at SNR shocks}

If SNRs are responsible for the bulk of GCRs, they should be capable of accelerating particles up to $\sim 10^{15}~\mathrm{eV}$. 
To test the hypothesis that the SNRs are the sources of GCRs, 
one must measure the maximum acceleration energy achievable at SNR shocks. 
X-rays, so far, are the most suitable energy range to probe accelerated particles near the maximum acceleration energies.  
In principle, TeV $\gamma$-rays can tell us the same information. 
However, current TeV data cannot reveal a clear signature of a spectral cutoff except for the brightest case, \rxj. 
Moreover, if TeV emission is due to inverse Compton (IC) scattering of relativistic electrons, 
cutoff energies can be determined by the Klein-Nishina suppression rather than due to 
a steepening of the  parent electron spectra.

The diffusion coefficient $D(E)$, which should be the same for electrons and protons, 
is necessary for calculating the energy dependent acceleration rate and 
the maximum attainable energies within the DSA theory. 
The shape of the energy spectrum of the accelerated electrons 
in the cutoff region depends only on $D(E)$, provided
that the acceleration proceeds in the synchrotron dominated
regime, where synchrotron cooling dominates over escape \citep{ZA07}.
Therefore, 
the wide band coverage of \astroh\ offers an interesting opportunity to probe 
$D(E)$ through the precise measurement of 
the spectral shape of synchrotron X-ray emission produced by accelerated electrons.

\subsubsection{Key question 3: Magnetic field amplification}\label{sec:MFA}

Magnetic field amplification (MFA), where the ambient magnetic fields are  enhanced  at a shock 
by a factor much larger than simple shock compression, 
 is now widely accepted as the key ingredient of 
 nonlinear DSA theory  \citep[e.g,][]{Bykov11_MFA_Review}. 
 MFA is thought to be driven by accelerated CRs via plasma instabilities. 
It has been proposed that 
diffusive CR current can non-resonantly excite strong magnetic turbulence in the precursor of a SNR shock \citep{Bell04}.  MFA is likely coupled with the acceleration efficiency 
because it would be driven by CR-induced instabilities.

High angular resolution observations of young SNRs with \chandra\ 
have revealed the presence of synchrotron filaments in young SNRs 
\citep{VL03,UAT03,Bamba03,Bamba05}. 
The magnetic field strength is estimated as $\sim 0.1$\,mG assuming 
that the filament widths are determined by rapid synchrotron cooling in the postshock region \citep{Voelk05}. This indicates that 
shocks in young SNRs are able to amplify the interstellar magnetic field by large factors. 

An alternative explanation for the narrowness of the filaments is a fast magnetic field damping behind a shock \citep{Pohl05}. This scenario also implies  similarly strong magnetic fields at shock fronts. Evidence for the amplified magnetic field comes also from the year-scale time variability of synchrotron X-ray filaments \citep{Uchi07,UA08}. If the variability timescale represents the synchrotron cooling time, the local magnetic field strength can be estimated to be as large as $\sim 1$\,mG. On the other hand, if the variability is due to intermittent turbulent magnetic fields \citep{Bykov09}, 
the time variability can be reconciled with a weaker magnetic field ($\sim 0.1$\,mG or less). A deep \chandra\ map of Tycho's SNR has revealed an interesting spatial feature, ``stripes'', of synchrotron X-ray emission, which may be taken as signatures of MFA and associated acceleration of CR protons and nuclei up to $\sim 10^{15}$\,eV \citep{Eriksen11}.

Unfortunately,  the angular resolution of \astroh\  will not be sufficient  to resolve 
the synchrotron filaments.  On the other hand, 
if the distribution of magnetic field strength in 
the synchrotron filaments contain a component of magnetic fields stronger than the average field, 
they would be particularly bright in hard X-rays. 
The hard X-ray measurement with \astroh\ may shed new light on the problem of MFA 
in young SNRs.

\subsection{Thermal X-ray emission from synchrotron-dominated SNRs}
\label{SNR:thermal_search} 

SNRs RX~J1713.7$-$3946 and RX~J0852.0$-$4622 (a.k.a.\ Vela Jr.) draw keen attention as 
cosmic-ray accelerators. 
The two young SNRs emit particularly bright synchrotron X-rays and TeV $\gamma$-rays. 
Peculiar characteristics of these SNRs are that their X-ray spectra are totally 
dominated by nonthermal components and no sign of thermal emission is 
found so far \citep{Slane_RXJ1713,Tanaka08}. 
While the $\gamma$-ray emission of most (if not all) SNRs measured with the \emph{Fermi} satellite is widely interpreted as being due to the decay of $\pi$-mesons, 
the well-studied SNR RX~J1713.7$-$3946, and also Vela Jr., 
are still matter of active debate \citep{ZA10,Ellison12,Fukui2012}.
Settling the dominant $\gamma$-ray emission mechanism in synchrotron-dominated SNRs 
leads to answer fundamental properties of the underlying particle acceleration mechanism, 
and the key questions discussed in Section~\ref{SNR:Intro}.

\begin{figure}[ht] 
\includegraphics[width=8cm]{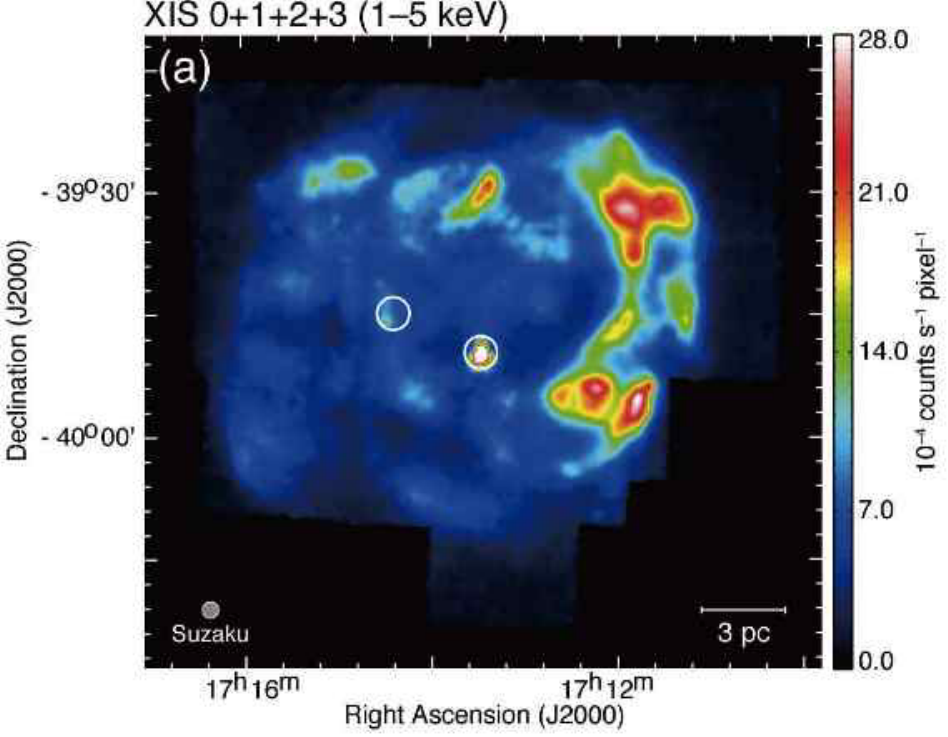}
\includegraphics[width=7.5cm]{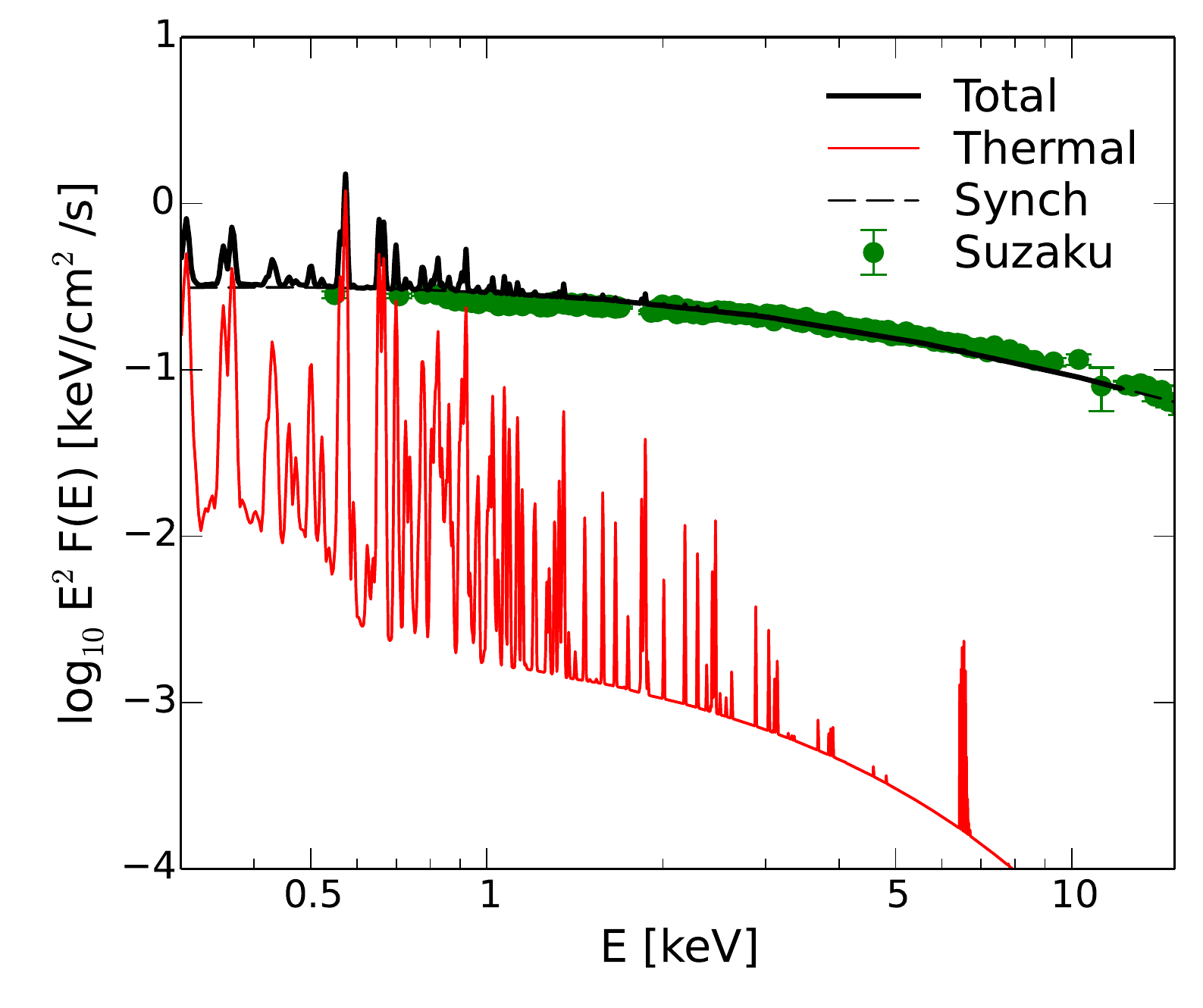}
\caption{(Left) 
\suzaku\ image (1--5 keV) of SNR \rxj\ \citep{Sano13}, where 
the X-ray emission is completely dominated by  synchrotron radiation. 
(Right) A predicted X-ray spectrum of SNR \rxj\ and the measured X-ray spectrum 
with \suzaku. The prediction shown in this plot is Model I of 
\citet{Lee_2012} (private communication with S.-H. Lee). } 
\label{fig:SuzakuRXJ}
\end{figure}

If one considers $\pi^0$-decay as the emission mechanism, a key parameter is the 
target gas density, 
since the flux of \pizero\ $\gamma$-rays is determined through convolution of 
 the target gas density and the CR proton density. 
One way of estimating the gas density is to measure thermal X-ray emissions. 
The density of shock-heated gas can be calculated from the volume emission measure (${\rm EM} \equiv \int n_e n_{\mathrm H} dV$) 
of the thermal component. 
Under a simple assumption that the shell is expanding into uniform gas \citep{Ellison10}, 
the upper limit on EM set by \suzaku\ suggests that IC scattering should be 
the major mechanism for the $\gamma$-rays\footnote{
Note that there would be some complications. 
For example, it may be the case that the shock of RX~J1713.7$-$3946 is expanding 
in a cavity created by the progenitor's wind \citep{berezhko2010,Ellison12,Fukui2012}. In this case, a significant amount of gamma rays might be emitted by protons/ions escaping from the SNR shell and then interacting with the dense cavity wall. 
The thermal X-rays, on the other hand, are emitted from low density gas in the cavity, and not related directly to the $\gamma$-rays.}.
Thermal X-rays, if detected, provide more direct information about the emission mechanisms for $\gamma$-rays. 

The \astroh\ SXS will allow  sensitive searches for weak thermal emission in 
synchrotron-dominated SNRs. 
The \astroh\ White Paper on Young SNRs addresses this issue. 
 The scientific goals include 
 elucidating the origin of the $\gamma$-ray emission, and thereby estimating the fraction of the explosion energy released in accelerated particles through non-linear shocks.
If the thermal X-ray lines were measured, 
the electron temperature determined from line diagnostics 
could be used to infer the fraction of shock energy consumed for heating postshock plasma under some assumption about electron heating. 
It has been shown for some young SNRs that  electron temperature \citep{Hughes_1E0102} or ion temperature \citep{Helder2009} is  lower than what is predicted from the Rankine-Huguniot relation.
These would be  taken as evidence of  ``CR-modified shock", where a significant amount of the shock energy is poured into cosmic-ray acceleration. 
 It is of interest to investigate whether reduced shock-heating can be seen in 
synchrotron-dominated SNRs such as RX~J1713.7$-$3946 and Vela Jr.

\subsection{Synchrotron X-ray spectrum beyond the cutoff} \label{sec:syncspec}

\begin{figure}[ht] 
\includegraphics[width=7.5cm]{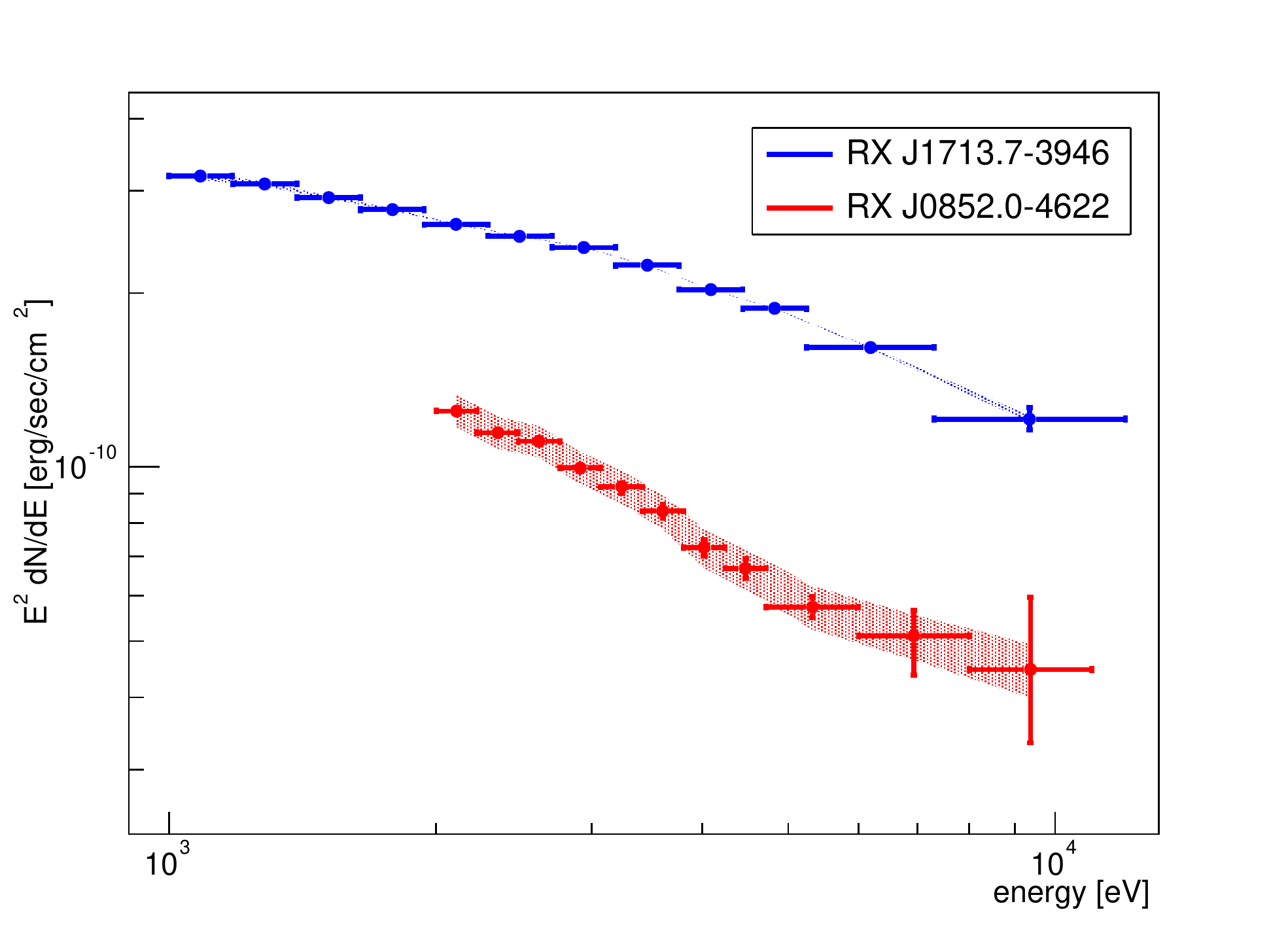} 
\includegraphics[width=7.5cm]{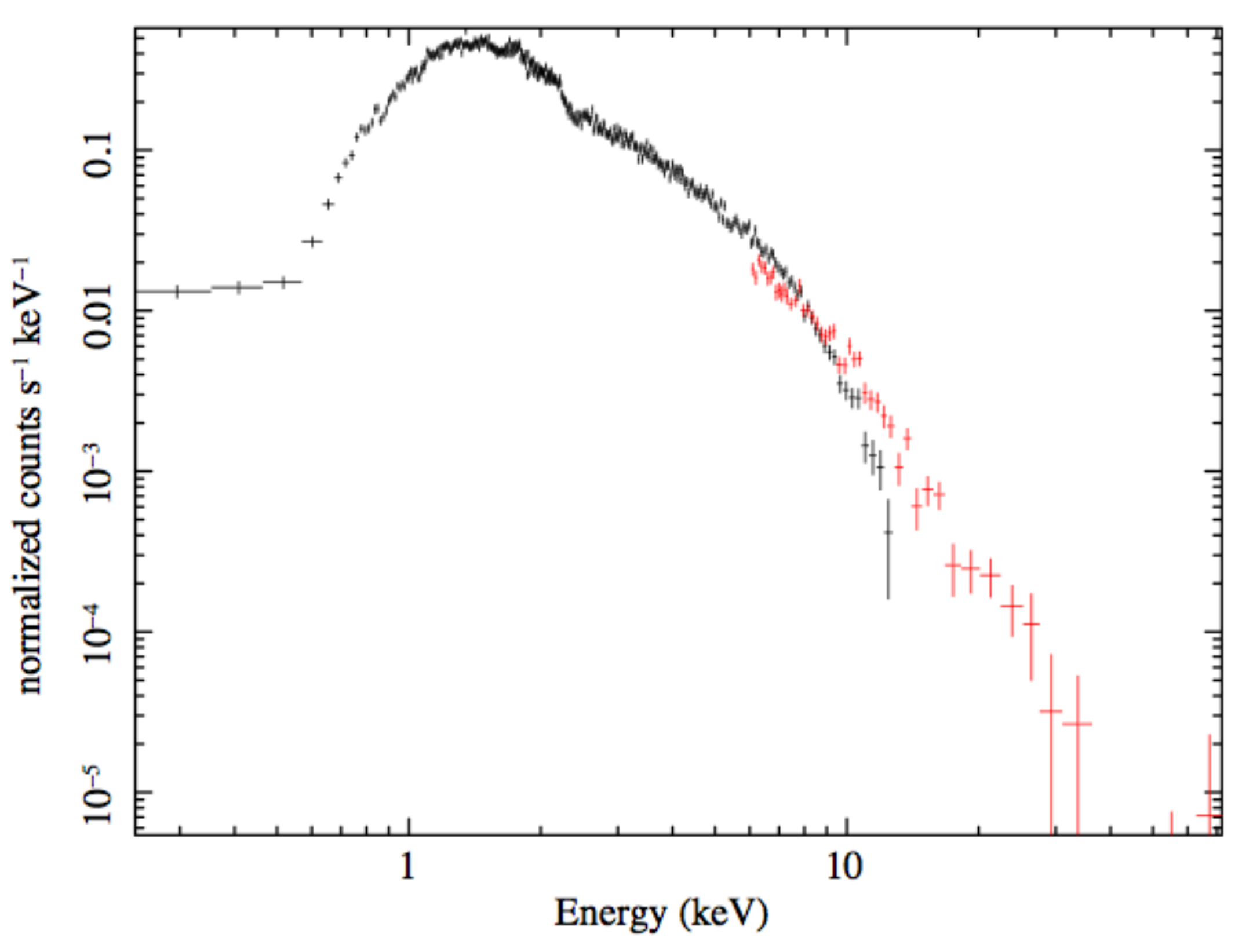} 
\caption{(Left) Spatially-integrated X-ray spectra of SNRs \rxj\ and Vela Jr.\ measured with \suzaku. 
Shaded regions represent systematic errors.  
(Right) Simulated SXI (black) and HXI (red) spectra of SNR \rxj\ (100 ks). The spectra 
are extracted from a $8\arcmin \times 8\arcmin$ box in the northwest rim. 
The simulation uses the Zirakashvili-Aharonian spectral form 
with $\epsilon_0 = 0.67$ keV \citep{Tanaka08}. }
\label{fig:RXJspec}
\end{figure}

As discussed in Section~\ref{SNR:Intro}, 
precise measurements of the spectral shapes of synchrotron peaks
can be used to constrain the diffusion coefficient at SNR shocks. 
 \astroh\ HXI measurements beyond the maximum would  confirm the Bohm diffusion or deviation from it using the Zirakashvili-Aharonian spectral form \citep{ZA07}.  
 Moreover, the sensitive hard X-ray observations 
above $\sim 20$ keV with the HXI may allow us to search for new spectral components 
besides the standard synchrotron component, such as nonthermal bremsstrahlung, especially in Cas~A \citep{Vink_brems}, the synchrotron radiation produced by 
secondary electrons and positrons, and jitter radiation. 

Again, we highlight synchrotron-dominated SNRs, \rxj\ and Vela Jr.,  which are the strongest nonthermal X-ray emitters among galactic SNRs 
(if integrated over the entire remnant), and therefore excellent targets to 
investigate the spectral shape of the synchrotron X-ray emission in great details. 
In Figure~\ref{fig:RXJspec}, we show spatially-integrated X-ray spectra of SNRs \rxj\ and Vela Jr.\ measured with \suzaku.  
While the X-ray spectrum of \rxj\ measured with \suzaku\ cannot be 
fitted with a simple power law \citep{Takahashi_RXJ_2008}, the 2--10 keV spectrum of 
Vela Jr.\ is consistent with a power law. A broadband X-ray spectroscopy is indispensable 
for placing strong constraints on the energy-dependent diffusion coefficient at shock waves. 

Figure~\ref{fig:RXJspec} shows simulated 
SXI and HXI spectra of SNR \rxj\ for an exposure time of 100 ks. 
The pointing position is assumed to be the brightest part of the northwestern rim.
The spectra 
are extracted from the $8\arcmin \times 8\arcmin$ HXI field-of-view. 
The simulation uses  the Zirakashvili-Aharonian synchrotron spectrum 
with a cutoff energy of $\epsilon_0 = 0.67$ keV \citep{Tanaka08}. 
Unlike the \suzaku\ HXD observations of \rxj\ \citep{Tanaka08}, \astroh\ observations will realize 
spatially-resolved hard X-ray spectroscopy. 
\astroh\ will provide important new information about the key parameters of 
the DSA theory and potentially also provide solid proof of the $\gamma$-ray 
emission mechanism. 
We can use the broadband measurement of the X-ray spectral shape to infer  
$D(E)$, which in turn can be used to calculate the spectrum of high-energy protons, 
assuming full confinement, and also the $\pi^0$-decay spectrum produced by 
these protons. 
Comparison with the TeV $\gamma$-ray data tests the hadronic scenario and 
furthermore it will provide insights into the possible 
effect of particle escape on the proton spectrum.

\section{Extreme particle acceleration in gamma-ray binaries and microquasars}

\subsection{Introduction}
Gamma-ray binary systems are a relatively new class of astrophysical object, currently consisting just of five firmly identified
representatives: \psrb \citep{HESS05_PSRB1259}, \ls \citep{HESS06_LS5039}, \lsi \citep{albert06-lsi}, 
HESS J0632$+$057 \citep{aharon06-innerGalaxy}, and \fgl \citep{1FGLJ1018_fermi,hess12}. The key feature of these binary systems is that their
bolometric radiation (with subtracted contribution from the optical companion) is dominated by emission produced in the gamma-ray
energy band above $1$~MeV. Typically, the gamma-ray emission varies with orbital phase, and sometimes very strong/rapid
changes of the gamma-ray radiation intensity, i.e., some type of periodic flares, are observed. Some gamma-ray binary systems
display a strictly periodic gamma-ray signal. In the case of \ls it was possible to obtain the orbital period by using the
data collected only in the VHE domain. Certain orbit-to-orbit variability has also been observed in other systems. In particular,
the high energy (HE) emission observed from \lsi with \lat displayed a clear change in 2009 March \citep{htt12}. 

All the gamma-ray binary systems should accelerate high energy particles with high efficiency. However, the implications
of these observations on acceleration theory is still missing, since gamma-ray binary systems display a very complex phenomenology
that includes presence of several radiation components with ambiguous relations, unusual spectral behavior, and high level of
orbital variability. Moreover, these objects demonstrate very individual properties that apparently cannot be explained by a
change of the basic parameters (separation distance, viewing angle, etc) within some general scenario.  Finally, the nature of the
compact objects in the gamma-ray binary systems have not yet been identified, except one clear case of \psrb, which consists of a
late-type star and a pulsar. Thus, gamma-ray binary systems remains a highly unexplored group of high energy 
sources, and their study with \astroh will allow to answer several important questions related to general properties of
nonthermal acceleration mechanisms and the specific processes taking place in these enigmatic sources.

Flaring HE or VHE gamma-ray emission components have also been detected from the X-ray binaries \cygxI \citep{cygx1} and \cygxIII
\citep{LAT09_CygX3}, but this emission is neither dominant nor coherent with certain orbital phases as is the case of the true
gamma-ray binaries. These observations however indicates that jets launched in conventional X-ray binary systems can also be
very efficient accelerators of  nonthermal particles.

\subsubsection{Key question 1: Study of the nonthermal acceleration process}

The concept of acceleration efficiency can be interpreted in different ways: either indicating a high maximal value of energy of
individual particles achieved under given conditions, or as a measure of the fraction of energy transferred to the population of
nonthermal particles.  It is very remarkable that gamma-ray binary systems seem to accelerate particles with unprecedented
efficiency {\em in both senses}.  The first type of efficiency is the best illustrated by a simultaneous study of X-ray and VHE
gamma-ray emission detected from \ls. It has been shown that in this system the effective accelerating field should contribute at
least at a level of $30\%$ of the total electromagnetic field \citep{Takahashi09}.  Binary pulsar system \psrb
provides us with an example of a remarkable conversion of the available energy to gamma-ray radiation. During the GeV flares that
occurred approximately one month after periastron passages in 2010 December and 2014 May, the gamma-ray luminosity measured with
\lat in the energy interval between $0.1$ and $1$~GeV exceeded $80\%$ of the spin-down luminosity of the pulsar
\citep{fermi11}. Importantly, this phenomena may have a counterpart in the X-ray energy band \citep{bzr14};  therefore study of this
phenomena with \astroh may shed light on its nature.

\begin{wrapfigure}{L}{7cm}
\includegraphics[width=7cm]{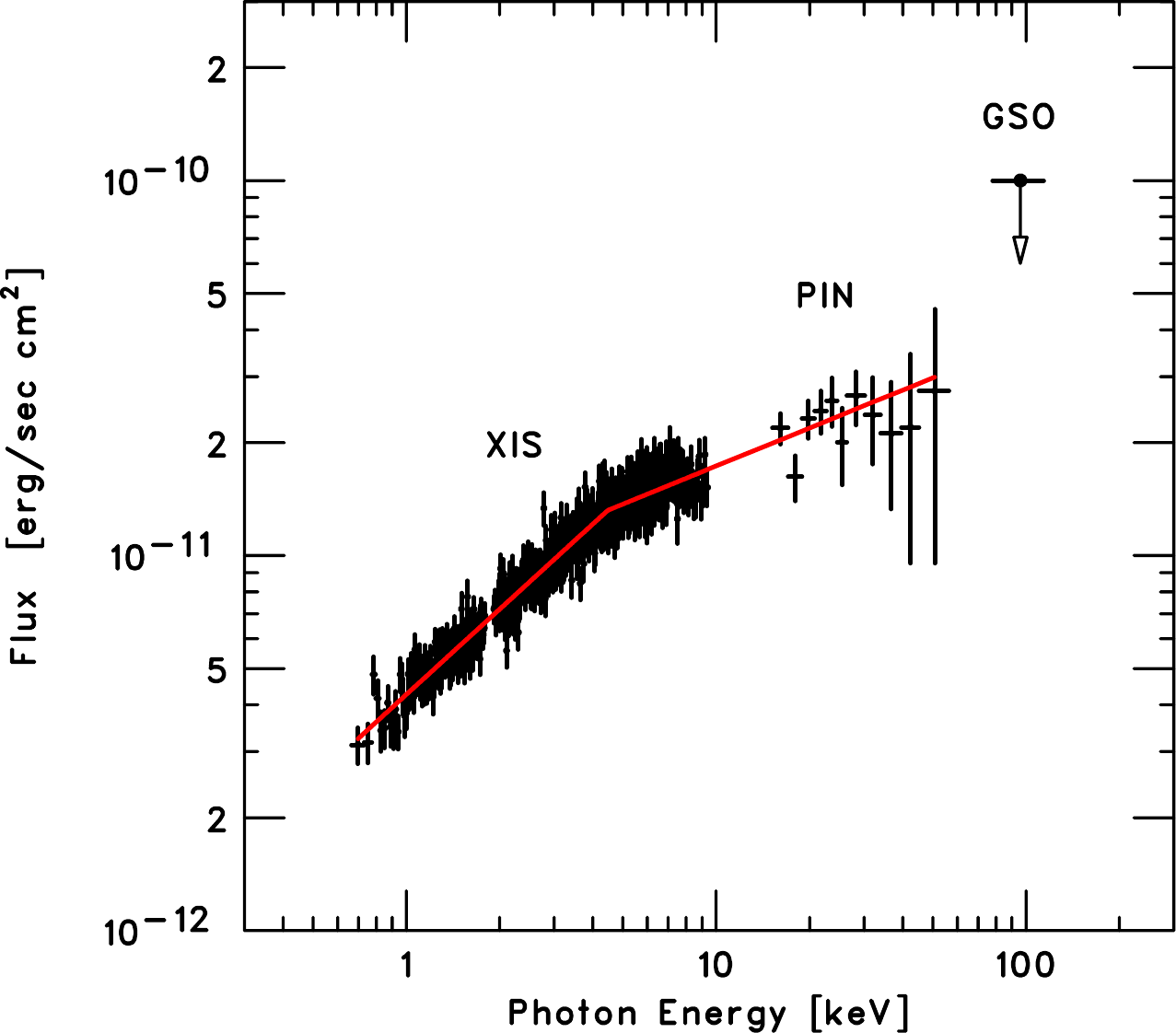}
\caption {\suzaku observations of the spectral break in \psrb.  Taken from \cite{uchiyama09}. }
\label{psrb_uchiyama}
\end{wrapfigure}

Although, currently only one gamma-ray binary system, \psrb, is confirmed to contain a pulsar, it was suggested \citep{Dubus:2006}
that the phenomena of gamma-ray binary systems are related to interaction of a non-accreting pulsar with the optical
companion. Therefore, these objects may have a very close connection to another (much more numerous) class of gamma-ray sources:
to pulsar wind nebulae (PWNe). Indeed, in a binary pulsar system containing a pulsar enough powerful to prevent accretion onto
neutron star (in contrast to accreting pulsar systems), one would expect a realization of the scenario of {\em compactified
  nebula} \citep[see e.g.][]{Maraschi.Treves:1981}. It means that all the components typical for regular PWN --
ultrarelativistic pulsar wind, pulsar wind termination shock, and shocked electron-positron outflow-- are present in such
systems. However, despite the same ingredients binary pulsar systems differ substantially from the plerions around isolated
pulsars. In particular this concerns the dynamics of the magnetic field in the shock downstream region. In the case of standard
pulsar wind nebula the magnetic field strength increases in the postshock region until magnetic field pressure reaches the
equilibrium with gas pressure \citep{Kennel.Coroniti:1984}. In a binary pulsar the magnetic field has a tendency of rapid decrease
of its strength towards flow downstream caused by bulk acceleration of the flow \citep{Bogovalov.Khangulyan.ea:2012}. This
property may have an important implication on the study of the nonthermal spectra formed at relativistic shocks. Indeed, in
isolated pulsars the bulk of the synchrotron emission is generated in regions of strong magnetic field, at distances significantly
exceeding the shock radius, $R_{\rm sh}$ (e.g., $\sim 10R_{\rm sh}$ in the case of the Crab pulsar). The electron spectrum present at
such distances does not correspond to the spectrum formed at the termination shock due to the impact of the cooling
processes. However, in the case of binary pulsars the strongest emission should be generated in the closest vicinity of the
termination shock, where the spectrum has not yet been deformed by cooling.  In particular, it was argued 
that the spectral break revealed with \suzaku in the X-ray spectrum of \psrb\ 
(Figure~\ref{psrb_uchiyama}) can be related to the lower energy cutoff of the
acceleration spectrum, associated with the Lorentz factor of the relativistic pulsar wind $\gamma \sim 4\times 10^5$ \citep{uchiyama09}.

\subsubsection{Key question 2: Compact object nature}

Proper implications of observations of gamma-ray binary systems in X- and gamma-ray energy bands are rather complicated, since the
physical scenario, that is realized in these systems, is not yet robustly understood. It is believed that the high energy activity
of these objects is governed by processes related to the compact object, which can be either a pulsar or a black hole. However,
the nature of the compact objects have not been identified in the majority of the cases. The pulsar is expected to release its
rotation energy via a cold ultrarelativistic pulsar wind, and in the black hole systems the main source of energy is
accretion. Thus, the properties of the acceleration sites, expected in these two scenarios, differ dramatically. In
the binary pulsar systems, the nonthermal particles are expected to be related to a strong
ultra-relativistic shock occupying a significant region in the system like in pulsar wind nebulae.  In contrast, jets launched by
the compact object most likely themselves are sites of particle acceleration.
 
A distinctive criterion of identification of an accretion powered system is related to the presence of thermal X-ray
emission. Such a component should be produced by the inner regions of the accretion disk. Currently, no thermal X-ray component
has been detected from gamma-ray binary systems. The lack of its detection cannot be however treated as conclusive evidence, since
there are several factors that can weaken it. For example, when the companion optical star does not fill the Roshe
lobe, the angular momentum of accreted matter is not high enough to form an accretion disk, thus such systems should lack bright
thermal X-ray emission. On the other hand, even in absence of accretion disk, these systems are still able to
launch a jet powerful enough to satisfy the energy budget required to explain the gamma-ray emission detected, e.g., from \ls \citep{Barkov2012}.

A possible way of determining the nature of the compact object is to study emission produced by the stellar wind. Indeed,
while in the case of direct wind accretion the stellar wind is not expected to be strongly perturbed by the compact object and the
jets, the presence of a powerful non-accreting pulsar in a binary system should strongly affect the stellar wind
\citep{Bogovalov.Khangulyan.ea:2008,khan2008,Bogovalov.Khangulyan.ea:2012}. In particular, a significant part of the stellar wind
is to be shocked and heated up to keV temperatures \citep{Zabalza.Bosch-Ramon.ea:2011}, and consequently thermal X-ray lines can
be studied with the micro-calorimeter onboard \astroh.  Detection (or equally non-detection) of this, currently not seen emission
component, should shed light on the nature of  the compact objects.

Finally, one can argue that compact binary systems harboring a non-accreting pulsar are not capable of producing  detectable pulsed radio signal
because of severe  absorption in the dense stellar wind. Moreover, uncertainties in definition of the orbital parameters do not
allow to search for pulsed gamma-ray component in \lat data collected from gamma-ray binary systems  \citep{ctr12}. Therefore, the
X-ray band remains the only plausible channel for detection of pulsed emission from these objects.

\subsubsection{Key question 3: Nonthermal activity in Galactic jets}

Generally, X-ray binaries are treated as {\em thermal} sources effectively transforming the gravitational energy of the compact
object (a neutron star or a black hole) into thermal X-ray emission radiated away by the hot accretion plasma.  However, since the
discovery of compact galactic sources with relativistic jets (dubbed as microquasars) the general view on the role of nonthermal
processes in X-ray binaries has significantly changed.  It is now recognized that nonthermal processes do play a non-negligible
role in these accretion-driven objects, and this nonthermal activity may result in acceleration of particles to HE and VHE
energies. \astroh can significantly contribute to study of the nonthermal processes in microquasars thanks to its high
sensitivity in the hard X-ray energy band and polarimetric capabilities of SGD in the soft gamma-ray energy band. This may shed light on
several fundamental processes, e.g., disk-jet connection, that are very important in jet sources on different scales.

\subsection{Extreme Particle Acceleration in Gamma-ray Binaries} 

As it was shown via interpretation of the multiwavelength data of \ls\ obtained with \hess and \suzaku, the efficiency of the acceleration process operating in this source should be extremely high. Namely, if the acceleration timescale is expressed as: 
\begin{equation} \label{eq:atime}
t_{\rm acc} = \frac{\eta R_{\rm L}}{c} \sim 0.1\,  \eta \left(\frac{E_{\rm e}}{\rm 1\, TeV}\right)  
\left( \frac{B}{\rm 1\, \ G} \right)^{-1} \ \rm s, 
\end{equation} 
where $\eta \geq 1$ parametrizes the acceleration efficiency, $B$ magnetic field strength and $E_{\rm e}$ electron energy, an
efficiency of $\eta < 3$ is required \citep{Takahashi09}. We note that the value of $\eta =1$ corresponds to the extreme
accelerators with the maximum possible rate allowed in magnetohydrodynamic setup. The accelerator operating in \ls\ might be one
of the most efficient in all astrophysical sources. Therefore the nature of the compact object remaining behind this acceleration
process deserves a special attention and may give an important insight into the physics of highly efficient accelerators.

The X-ray emission allows a deeper insight on the properties of the highest energy particles residing in gamma-ray binary
systems. Obviously, these particles predominately emit in the $\gamma$-ray energy band, but the presence of the dense photon field
provided by the companion star almost unavoidably leads to a severe $\gamma$-$\gamma$ attenuation of TeV $\gamma$-rays
\citep{khan2008,Dubus.Cerutti.ea:2008}. Therefore, the intrinsic $\gamma$-ray spectrum can be significantly deformed. Thus, the
understanding of the very high energy processes occurring in gamma-ray binary systems requires a simultaneous study of the many
different processes, which include particle acceleration, transport, radiation, attenuation and eventually cascading. On the
other hand, the nature of $\gamma$-ray binary systems strongly favors the leptonic origin of the VHE emission, i.e.\ through IC scattering of VHE electrons on the stellar photons. In the frameworks of this scenario, the electrons responsible
for $\gamma$-ray emission should also emit in the X-ray band through the synchrotron channel. This emission remains insensitive to
the severe internal $\gamma$-$\gamma$ absorption.  Thus, sensitive X-ray observations together with TeV data provide a mean for a
deep insight into the physics of the highest energy particle population in TeV binary systems. However, current X-ray telescopes
lack both required sensitivity in the soft $\gamma$-ray and angular resolution in hard X-ray bands.

In compact binary systems the photon field is expected to be very intense. However in the VHE regime the acceleration process is
typically limited by the synchrotron losses, since the IC losses are severely suppressed by the Klein-Nishina effect.
In this case if the acceleration time is determined by Eq.~\eqref{eq:atime}, the synchrotron maximum energy appears to be
independent on the strength of the magnetic field. The only parameter that alters this energy is the acceleration efficiency $\eta$:
\begin{equation}\label{eq:syn_max}
\hbar\omega_{\rm max}\simeq 150\,  \eta^{-1}\  \rm MeV.
\end{equation}
For the value of $\eta\simeq3$ inferred for \ls, the synchrotron spectrum should extend to energies $\hbar\omega_{\rm
  max}\sim50\rm \,MeV$ \citep{Takahashi09}. This prediction seems to be consistent with orbital phase dependence of gamma-ray
emission detected with \comptel from this source \citep{collmar}. Moreover, the flux level obtained with \comptel is consistent
with extrapolation of the X-ray spectrum obtained with \suzaku. Thus, if this interpretation is correct, one should expect a
 powerlaw spectrum in an energy interval unprecedentedly broad for binary systems: from $\sim1$~keV to $\sim30$~MeV. The luminosity emitted in this
radiation component should allow a very detailed study of it with \astroh, which would include precise measurements of variability
and spectral features. In particular, if the binary pulsar scenario is realized in \ls, one may expect a detection of a low energy break that is related to the bulk Lorentz factor of the pulsar wind; such a feature has been suggested in the \suzaku data obtain from \psrb \citep{uchiyama09}.

\begin{wrapfigure}{L}{8cm}
\includegraphics[angle=270, width=8cm]{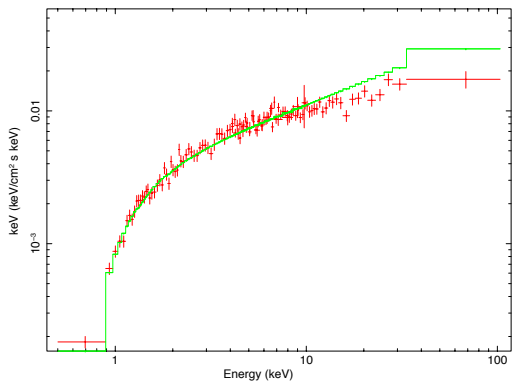}
\caption {Simulation of the \astroh\ HXI observation of \psrb\ with an exposure of 10 ks
 as it crosses the disk.  The spectral parameters are derived from the \suzaku\ observation shown in Figure~\ref{psrb_uchiyama}. Spectral data are fitted with a single power law.}
\label{psrb_hxi}
\end{wrapfigure} 

\psrb is the only gamma-ray binary with a well identified compact object; it is a 48 ms radio pulsar in a highly eccentric 3.4
year orbit with a Be star LS 2883.  \suzaku observations in 2007 for the first time indicated a break of the spectrum
\citep[Figure~\ref{psrb_uchiyama};][]{uchiyama09}.  The observed X-ray spectral break was attributed to the low-energy cutoff of the
synchrotron radiation associated with the Lorentz factor of the relativistic pulsar wind $\gamma \sim 4\times 10^5$. Thus, this
break gives us additional important information on the energy of the shocked relativistic electrons and the magnetic field.
It should be noted that the presence of the nearby X-ray pulsar IGR J13020-6359, located only 
$10\arcmin$ from the \psrb\ makes the results of the \suzaku HXD  model-dependent. 
Therefore, hard X-ray imaging observations  would be a   great benefit. 
Figure~\ref{psrb_hxi} shows a simulation of 10 ks observation of \psrb\ with 
\astroh, where  we show that if one would try to fit the broken power-law spectrum that was revealed by \suzaku with a single power-law model, the high energy data will clearly deviate from the model. 

Importantly, if this spectral feature is real, it should be much more prominent in binary pulsar systems than in PWNe around isolated pulsars. Cooling process together with particle transport in strengthening  magnetic field  should dilute this spectral break.

\subsection{Nature of Compact Objects in \ls}
In this subsection we focus on one example of gamma-ray binary system \ls that contains yet an unidentified compact object. \ls is the only persistent gamma-ray binary with an O-type star companion. In this system a compact object is in a
mildly eccentric 3.9 day orbit around O6.5V(f) main star \citep{casares2005_ls}. The source shows periodic X-ray emission with a
peak around the inferior conjunction \citep{Takahashi09} and  a modulation practically mirrored by its VHE emission
\citep{HESS06_LS5039}. \lat observations showed that, similar to TeV and X-rays, GeV emission is modulated with the orbital
period, with flux variations of a factor three along the orbit \citep{LAT09_LSI}.  The spectral shape at the GeV range is
consistent with a power law with an exponential cutoff at energies of a few GeV (with the position of the spectral cutoff energy
stable along the orbit). The combined effects of adiabatic losses, synchrotron and IC cooling, and anisotropic IC emission
applied to a single particle population with a power-law energy distribution are unable to reproduce a cutoff at a few GeV while
accounting for the hard spectrum from 10 GeV to 1 TeV \citep{khan2008}.

\citet{Zabalza2012} have proposed a model that assumes two different locations for the production of the observed GeV and TeV components of the $\gamma$-ray emission. The apex of the contact discontinuity was proposed as the candidate location for the GeV emitter,
and a pulsar wind termination shock in the direction opposite of the star, which appears because of the orbital motion, as the
candidate location for the TeV emitter. \citet{Barkov2012} proposed an alternative model, which considers the wind accretion onto a rotating black hole in the close binary system harboring a young massive star. It was shown that the angular momentum of
the accreted stellar wind material is not sufficient for the formation of an accretion disk, but the powerful jets can still be
launched in the direction of the rotation axis of the black hole. In this case no observational signatures of accretion, as
typically seen from the thermal X-ray emission from the accretion disks, are expected in the suggested scenario. The obtained jet
luminosity can be responsible for the observed GeV radiation if one invokes Doppler boosting.

\begin{wrapfigure}{L}{7.5cm}
\includegraphics[width=7.5cm]{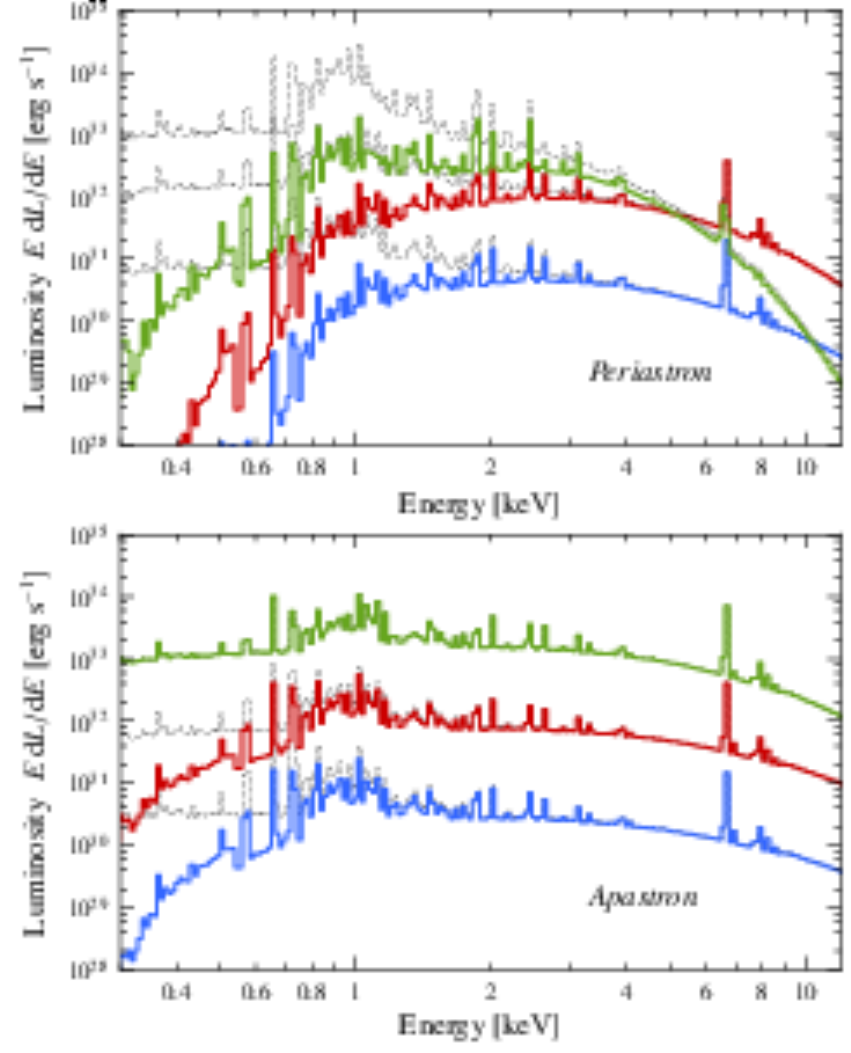}
\caption {Thermal X-ray emission expected from \ls. 
Taken from \citet{Zabalza.Bosch-Ramon.ea:2011}.}
\label{ls_zabalza}
\end{wrapfigure}

A possible observational test that allows to discriminate between these two scenarios is based on future observations with \astroh; if the binary pulsar scenario is realized, one should expect a detectable contribution from shocked stellar wind. In
Figure \ref{ls_zabalza} the computed spectrum of thermal X-ray emission is shown for three different values of the pulsar
spin-down luminosity (see \citet{Zabalza.Bosch-Ramon.ea:2011} for detail). The spectra correspond to the conditions expected in \ls at periastron and
apastron orbital phases. However, it should be taken into account that \ls is a bright source of nonthermal X-ray radiation. The most
detailed X-ray spectrum at different orbital phases was obtained with \suzaku in 0.7--70 keV energy range \citep{Takahashi09}.
The X-ray emission observed with \suzaku is characterized by a hard power law with a phase-dependent photon index which varies
within $\Gamma = 1.45\mbox{--}1.61$, moderate X-ray luminosity of $L_X \sim 4\times10^{33} \rm erg\ s^{-1}$. 
Since the luminosity of the nonthermal component exceeds the expected emission from the shocked wind, the thermal contribution
can be inferred only with usage of the \astroh\ SXS.
 If no line features will be detected with \astroh, that would imply a
rather strict upper-limit on the spin-down luminosity of the pulsar that closely approaches the lower-limit imposed by the energy
requirement to power the bright gamma-ray emission component.

Although the \suzaku observation revealed no spectral features in the broadband X-ray emission, it should be noted that since \ls is located close to the Galactic plane, the contribution from the Galactic ridge X-ray emission (GRXE) is not negligible. 
Since the \suzaku HXD is a non-imaging instrument authors had to assume a spectral shape of the GRXE  to subtract it from the observed spectrum. Thus \astroh observations are very important both to check the evolution of the X-ray emission and to check whether the broadband X-ray emission of \ls is indeed completely featureless, which is extremely important in order to understand the true nature of the source and to select between the proposed models. 
 
\subsection{Hard X-ray emission of the BH jets}\label{sec:microquasar} 

Approximately  20 per cent of the 
$\sim 250$  known X-ray binaries show  synchrotron radio emission, 
and observations in  recent years have revealed  the presence of radio 
jets in  several classes of X-ray binary sources. 
The high brightness temperature and the polarization of the  
radio emission from X-ray binaries are indicators of 
the  synchrotron origin of radiation. 
The nonthermal power of synchrotron jets
(in the form of accelerated electrons and kinetic energy of the 
relativistic outflow)  during strong radio flares could be comparable with,  
or even exceed,   the thermal X-ray luminosity of the central compact object. 

If the acceleration of electrons proceeds at  a very high rate,  the spectrum of 
synchrotron radiation of the jet can extend   to hard X-rays/soft $\gamma$-ray domain 
\citep{AtoyanAharonian99,Markoff01}.
In addition,  the  high density 
photon fields  supplied  by  the accretion disk   and  by  the  
companion star,  as well as  produced by the  jet itself,   create favorable conditions 
for effective  production of X- and $\gamma$-rays 
of IC origin inside the jet \citep{LevinsonBlandford96,AtoyanAharonian99,Georganopoulos02}.
Generally, this radiation is expected to have  an episodic character associated with strong radio  
flares in  objects like  GRS~1915+105.

The previous observations with OSSE and COMPTEL showed  that the spectra of  
microquasars, in particular GRS~1915+105 and  Cyg X-1, 
extend  to the domain of  hard X-rays and soft $\gamma$-rays.   
For any reasonable temperature of the  accretion plasma,  models of  thermal 
Comptonization  cannot explain the MeV radiation, even  when  one invokes the 
so-called bulk-motion  Comptonization.   For explanation of this 
excess,   the so-called ``hybrid thermal/nonthermal Comptonization'' model  has been 
proposed which assumes that the radiation consists of two 
components -- (i) the thermal Comptonization component  with a conventional  temperature of the 
accretion plasma $k T_{\rm e} \sim 20-30 \rm \  keV$  
and (ii) a nonthermal high energy component produced during the development 
of a linear pair cascade  initiated by relativistic particles in the accretion plasma 
surrounding the black hole \citep[for a review see][]{Coppi99}. 
This  model  requires  existence  in  the accretion plasma of a  relativistic 
electron population,  as a result of either direct electron acceleration or through 
pion-production processes  in the two-temperature accretion disk  with 
$T_{\rm i} \sim 10^{12} \ K$ \citep{Mahadevan97}. 

An alternative site for production of  hard X-rays and low energy $\gamma$-rays  could be the synchrotron jets.
In particular,  it has been proposed that the synchrotron emission of microquasars might extend to X-ray energies, either in the extended jet structure \citep{AtoyanAharonian99} or close to the base of the jet \citep{Markoff01}. Recently a significant contribution of  the nonthermal  X-ray emission  to the total X-ray luminosity of  Cyg X-3 has been argued  based on the detection of $\gamma$-rays by the \emph{Fermi} LAT and \agile \citep{Zdziarski12}.  The confirmation of the synchrotron X-ray component of jets in microquasars  not only will help to understand the acceleration mechanisms in these objects, but also add a key information to the disk-jet relationship.

The most promising energy band for the extraction of the synchrotron component is the hard X-ray to the soft $\gamma$-ray band where the radiation from the accretion plasma is suppressed.  
 \astroh\  has an excellent capability for detailed spectroscopic and  temporal studies of the  most prominent representatives of  microquasars  like GRS~1915+105, Cyg X-1, and Cyg X-3.  Detection  of polarization by the \astroh\ SGD  would  provide crucial test of the synchrotron origin of radiation. In this regard, one should mention the claim of the detection of polarization of hard X-ray emission above 400 keV  by \integr\ which can be explained only by synchrotron emission \citep{Laurent11}.

\section{Ultra high energy particles in galaxy clusters}

Accretion shocks in Clusters of galaxies are able to accelerate  protons  to extremely high 
energies.   The  energy distribution of  accelerated protons can be calculated 
self-consistently  via a  time-dependent numerical treatment of the nonrelativistic diffusive 
shock acceleration (DSA) process  versus energy losses caused by interactions with the comic microwave background  radiation (MBR).  In this scenario, 
the maximum energy of protons is achieved around $10^{19}$eV,  when the rate of DSA, 
determined by the  shock speed,  cannot overcome anymore the progressively growing  
lose rate due  to the Bethe-Heitler pair production processes.   The secondary  (``Bethe-Heitler") 
electrons and positrons  are immediately cooled down leading to a broad band  emission consisting of two, synchrotron and inverse Compton (IC) components. In the case of strong shocks, 
the substantial fraction of the available energy (10 \% or even more  if  the acceleration  takes place in a nonlinear regime)  is released in highest energy protons
from the cutoff region.  Furthermore,  the energy of these protons with 100 \%  efficiency is converted, via 
radiation of secondary  electrons, to nonthermal emission.   
Thus we deal with an extremely effective emitters of high energy emission.  Despite 
the remote locations of galaxy clusters  and  their  large angular extensions,  the hard X-ray  
emission initiated by ultra-high energy cosmic rays can be detected by  \astroh. 
This would open a new research area  of  studies  of extragalactic cosmic rays with 
\astroh. The synchrotron flux  peaks  around  100 keV to 1 MeV  
with a  very hard spectrum (photon index $\approx 1.5$)   below 100 keV.  
This spectral feature can be used  for an effective separation of this radiation component 
from the thermal X-ray emission of the hot intracluster plasma.
The detection of some nearby representatives of this source population, like the Perseus and  Coma clusters, in both X-ray and TeV bands should allow an independent and robust estimate of the intracluster magnetic field. Since $\gamma$-rays above 10 TeV are arriving with a significantly distorted spectrum due to  the energy-dependent absorption in the the Extragalactic Background Light (EBL), the comparison of the  hard X-ray and TeV 
$\gamma$-ray spectra  from clusters  should allow quite robust predictions  for the flux of EBL at mid infrared wavelengths.

\subsection{Nonthermal X-ray observations of galaxy clusters} 

The diffuse synchrotron radio emission detected from significant  fraction (about one third)  of rich clusters of galaxies  indicates the presence  of relativistic electrons in these largest structures in the Universe. Detections of nonthermal X-ray emission  have  been claimed  from a few galaxy clusters;  
generally it is interpreted as IC emission from the same population of electrons 
\citep[e.g.,][]{ff, eckert}.  However,  this interpretation requires a quite small magnetic field of the order of  $0.1 \mu$G, which is in stark contrast to the Faraday rotation measurements  which demand much 
stronger intracluster magnetic field of order of a few  $\mu$G  \citep{Clarke,Cluster_Bfield}. 
Several possible explanations of this controversy have been discussed in the literature. 
Two  most likely options in this regard  could be the simplest ones, namely  either the reports of detections 
of nonthermal X-rays are not correct,  or they are correct but the nonthermal 
X-rays are not of IC origin. 
Although there is a hope that \astroh\  will contribute  to the clarification  of this important 
issue, in particular  should be able to  
inspect the previous clams on the  nonthermal X-ray excess,   it is clear that if 
the  Faraday rotation measurements give a  realistic  estimate of the intracluster magnetic field, 
the chances will be quite  small for \astroh\  to detect IC X-rays even from the most promising 
clusters like Coma or Perseus.

\begin{figure}
\begin{center}
\includegraphics[width=6.5cm]{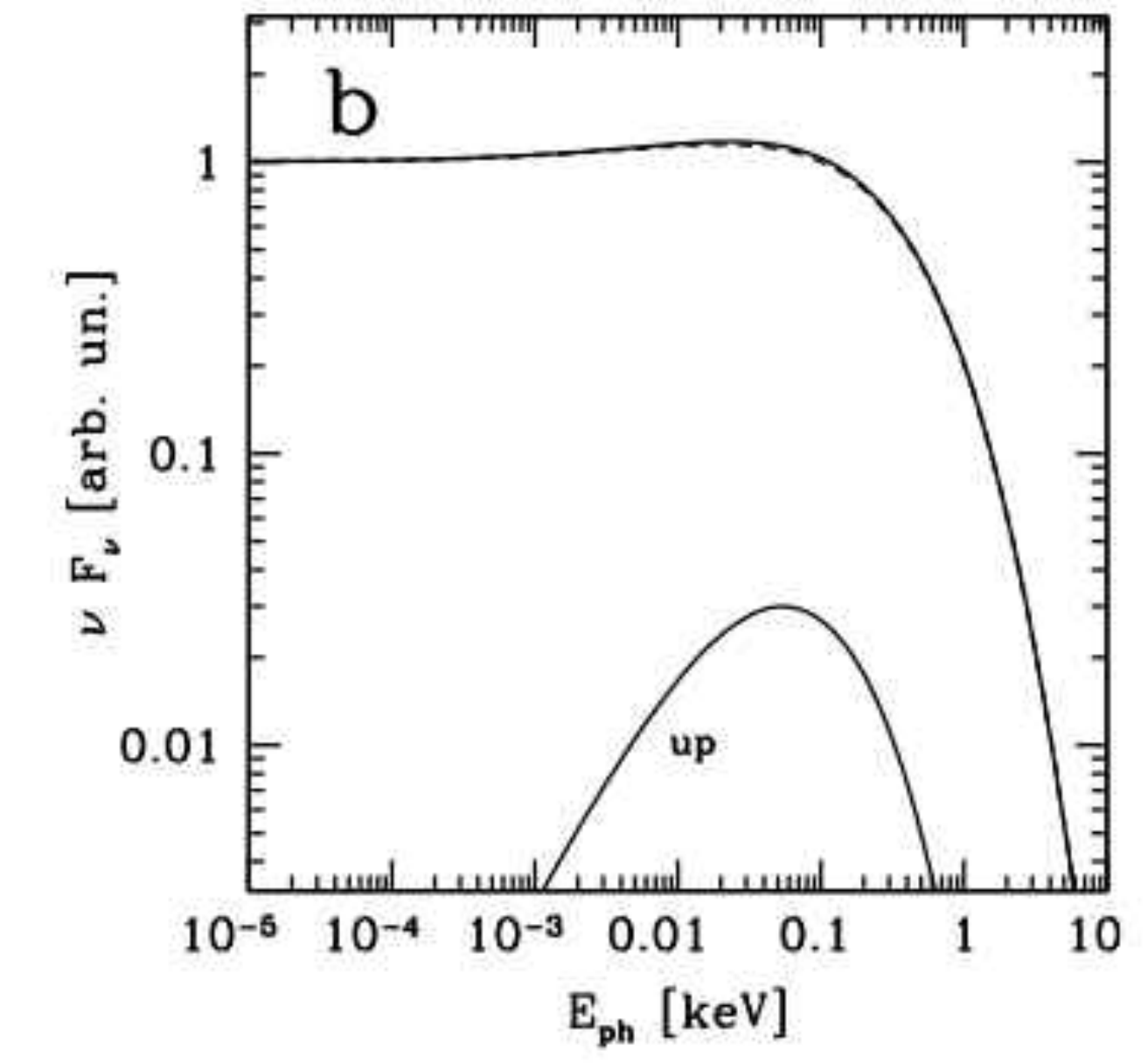}
\caption{
Spectral energy distribution of synchrotron radiation of electrons directly accelerated 
at the accretion shock front for a shock speed of 2000 $\rm km\ s^{-1}$ and the intracluster 
magnetic field $B_1=0.3~\mu$G.  It is assumed that the acceleration proceeds in the Bohm regime. The contribution from the upstream region (marked as ``up'') is  negligible compared to the contribution from the 
downstream because of lower density of electrons in upstream as well as due to the higher (compressed) magnetic field upstream.  Taken from \citet{gv1}. }
\label{fig:Cluster1}
\end{center}
\end{figure} 

On the other hand, we  may expect  detectable nonthermal X-rays  from another radiation channel, namely 
through  synchrotron radiation of extremely high energy electrons. While the radio synchrotron and
IC X-rays are produced by relatively modest (GeV energy)  electrons, the extension of 
synchrotron radiation to the X-ray band  requires  multi-TeV parent electrons.   
These electrons   cool  via synchrotron radiation and IC scattering   on very short  timescales, thus they 
cannot travel too far from the sites of their production. Therefore, in the case of shock acceleration we
should expect  very narrow filamentary structures  which would mimic the spatial structure of the shocks.

In principle, we can expect diffuse synchrotron radiation if the electron acceleration has 
stochastic (Fermi II type) character.  However, for any realistic parameters characterizing the intracluster medium it is difficult to accelerate electrons to TeV  energies, thus the synchrotron peak should appear 
well below 1 keV.  Since the  shock speeds in galaxy clusters  do not exceed a few thousand km/s, 
even in the  case of Bohm diffusion   the synchrotron peak appears  below 1 keV 
(see Figure~\ref{fig:Cluster1}).
Note that when the energy losses of electrons are dominated by the synchrotron cooling, and 
the acceleration proceeds in the Bohm diffusion region, the position of synchrotron peak does not depend
on the magnetic field.  For example, for the shock speed of a few 1000 $\rm km\ s^{-1}$ it is expected around 1~keV \citep{ZA07}.  Thus, for the typical  
infracluster magnetic fields of about  $B_1=0.3~\mu$G, electrons are predominantly cooled  via IC scattering, therefore 
the synchrotron peak  is shifted to  0.1~keV \citep{gv1}.  However, one 
should mention that in a highly turbulent 
medium, when the  the characteristic scale of $\lambda$ is less than the electron gyro radius, $r_{\rm g}=m_{\rm e} c^2/eB \sim 10^{11}  (B/1 \mu  \rm G)^{-1} \rm cm$,  the peak of radiation could be shifted towards higher energies by a factor of $k=r_{\rm g}/\lambda > 1$.

\subsection{Synchrotron radiation  of the Bethe-Heitler electron-positron pairs}

Clusters of galaxies  are  potential sites for effective acceleration of  cosmic rays 
\citep[see][for a review]{revste}. In particular, it has been argued that large scale accretion shocks can effectively accelerate electrons and protons up to ultrarelativistic energies \citep{norman,volk_cluster,ber_cluster,loeb,kushnir,miniati,blasimerger,steshocks,ryu}.
Formally, according to the so-called Hillas criterion, 
galaxy clusters are  amongst the few source populations capable, as long as this concerns the dimensions of the structure and the value of the magnetic field, to accelerate protons up to $10^{20}$ eV.

The interactions of ultra-high energy protons with the MBR lead to production of electrons with  
energies $E \geq 10^{15} \ \rm eV$ which is not  accessible through any direct acceleration mechanism. 
These electrons   cool  via synchrotron radiation and IC scattering   on very short  timescales 
(compared to  both the age of the source and  interaction timescales of protons); they cannot propagate 
too far from the sites of their production,  i.e. they are  localized in space. This implies that   
the corresponding radiation components in the X-ray and $\gamma$-ray energy bands 
are tracers of primary protons, and contain precise  information about the acceleration and propagation 
of their ``grandparents".  Since protons can freely travel within the cluster, we should expect diffuse X-ray emission from regions with linear size in excess of a few Mpc. 

It is remarkable that as long as   the maximum energy of protons is determined by radiative losses,  one can ignore the escape of the highest energy particles  many aspects of which remain not well understood and contain large uncertainties. This allows  good accuracy of calculations of the  spectra of accelerated particles, 
their secondary products, and,  consequently, robust predictions for the synchrotron radiation of the secondary electrons.  The detection of this radiation is possible only if the acceleration proceeds  close to the Bohm diffusion. In this case three key model parameters determine the conditions for effective production of the secondary synchrotron radiation; these are   the  fraction of available energy converted to accelerated particles,  $\kappa_{\rm CR}$, the shock speed $v_{\rm shock}$, and the  average magnetic field, $B$.

\begin{wrapfigure}{r}{7cm}
\includegraphics[width=7cm]{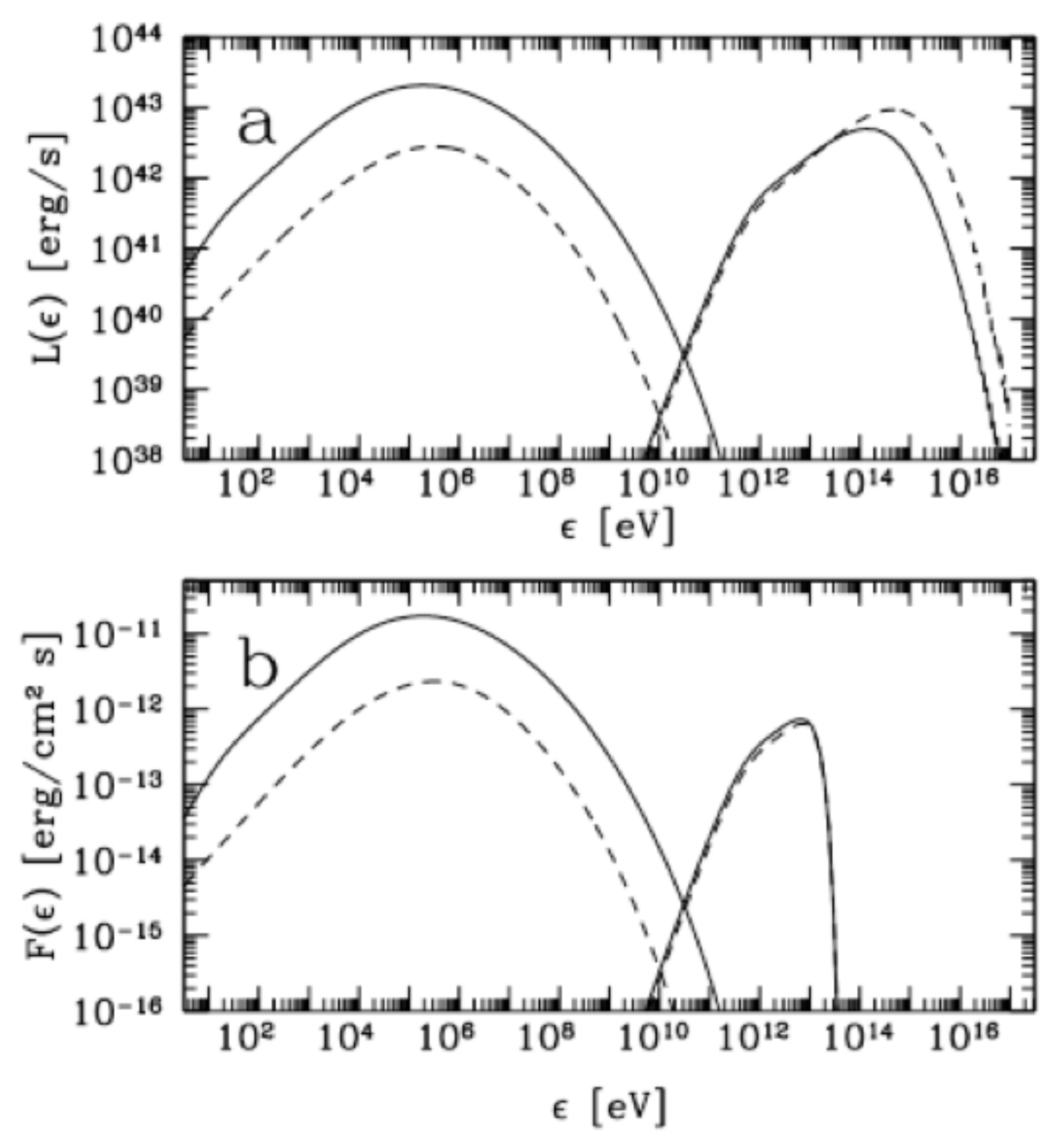}
\caption{
(a) The luminosity of the broadband synchrotron and IC radiation components from downstream (solid line) and upstream (dashed line). (b) Energy flux at the observer location, after intergalactic absorption of gamma-rays using the EBL model of \citet{ebl} for a distance to the source 100 Mpc. Taken from \citet{gv2}.}
\label{fig:Cluster2}
\end{wrapfigure}

Proton acceleration in galaxy clusters  by accretion shocks has been  studied in \citet{gv2}.  For realistic  shock speeds of a few thousand km/s and a background magnetic field close to  $1\, \mu  \rm G$, the maximum energy achievable by protons is determined by the energy losses  due to  Bethe-Heitler pair production;  it ranges from $10^{18}$ eV to $10^{19}$ eV \citep{gv2}. 
In this scenario, particle  acceleration operates on timescales comparable to the age of the cluster, $t \sim 10^{10}$yr. However,  since in such sources  the  steady state is never achieved, a time-dependent treatment is required. The proton energy  spectra  exhibit interesting features. 
Because of the specific energy-dependence of energy losses  caused by  the Bethe-Heitler pair production, the decay of the  proton spectrum around  the cut-off energy appears quite different  from the usually assumed exponential shape.  Another interesting effect is related to the specifics of energy losses of electrons due to IC scattering which  proceeds in the Klein-Nishina  regime.  These effects leads to the formation of broad-band radiation of electrons consisting of the synchrotron and IC components. The curves shown in Figure~\ref{fig:Cluster2} 
are obtained for acceleration by nonlinear shocks modified by the pressure of relativistic particles, 
and assuming for the compression factor $R=7$. The synchrotron spectrum in this case before the cutoff is very hard with a spectral  index $\alpha=1.5$ \citep{MD01}. Note that this type of spectrum provides  significantly more effective release of energy of accelerated  particles to high energy radiation compared to the  $E^{-2}$ type proton  spectrum (formed in the case of linear shock acceleration) because the major fraction of the nonthermal energy is accumulated in the cutoff region.

The  detectability of clusters in hard X-rays and $\gamma$-rays  associated with  interactions of ultrahigh energy protons with the MBR photons,  linearly depends on the value of the parameter $A=W_{62}/d_{100}^2$, where $W_{62}=W/10^{62} \ \rm erg$ is the total energy released in accelerated particles,  and  $d_{100}=d/100$ Mpc is the distance to the source.
The luminosities and fluxes of the synchrotron and IC components shown in 
Figure~\ref{fig:Cluster2} are calculated for  $A=1$, assuming for 
the shock speed $v=2000\ \rm km\ s^{-1}$ 
and for the magnetic fields upstream $B_1=0.3$~G and downstream $B_2=7 B_1$. For the chosen parameters,  
the synchrotron and IC  luminosities  are similar.  Nevertheless,  while the  energy  
flux of hard X-rays at the distance of 100 Mpc from the source is expected at the level of 
$10^{-11} A$ erg cm$^{-2}$ s$^{-1}$, the  flux of IC $\gamma$-rays at multi-TeV energies is suppressed by an order of magnitude.  The reason is  the severe intergalactic absorption.  It is seen from Figure~\ref{fig:Cluster2}. 
that  although the maximum of the $\gamma$-ray luminosity  is located above 100 TeV,  the intergalactic absorption makes it invisible. 

Coma and Perseus clusters are two most promising objects in the nearby Universe (within 100~Mpc) to be detected in secondary hard X-rays and IC $\gamma$-rays. However, even for them the surface brightness is quite low given their large angular extensions;  this makes the detection of X-rays not an easy task. Indeed,  the accretion of  the cold external material onto a hot rich cluster of galaxies can lead to the formation of a strong shock at the position of the virial radius of the cluster. For Coma cluster the virial radius is about 3~Mpc, which for the distance to the source corresponds to the angular radius of about $1.8 ^\circ$. The angular virial radius of Perseus is somewhat less, $\approx 1.4^\circ$.  Thus, the angular surface of both objects is about  10 deg$^2$, and correspondingly 
the flux of hard X-rays  is about  $\approx 10^{-12} A \  \rm erg\ cm^{-2}\ s^{-1} \ deg^{-2}$.  The comparison of  this  flux estimate with the sensitivity limit  
of \astroh\ (for 100 ks exposure time )  of about  $5 \times 10^{-12} \ \rm erg\ cm^{-2}\ s^{-1} \ deg^{-2}$ at 20 keV,  gives us a quite robust  condition: 
$A \geq 5$, or $W \geq 5 \times 10^{62} \ \rm erg$  given that  both objects are located at a distance of about 100~Mpc.  Thus the  hard synchrotron X-rays  of secondary (``Bethe-Heitler") electrons 
can be detected only if  approximately 5  \%   of the total energy of these clusters ($\approx 10^{64}$erg) is transformed  to  ultra-high energy cosmic rays. Although this is a quite tough condition, it is in agreement with 
predictions of DSA. Moreover,   in the case of realization of nonlinear regime of DSA, the acceleration efficiency could be  as high as 50 \% \citep{MD01}. In this case, 
\astroh\ should be able to detect this component with high statistical significance. 

Although the integral flux of  X-rays decreases with the distance to the source as $1/d^2$,
for the fixed physical size of the source  (i.e. the  virial radius), the 
brightness distribution does not depend on the distance $d$.  Therefore,  as long as 
the detectability is determined by the surface brightness,  the remote powerful 
clusters of galaxies with energy budget $W \approx 10^{64}$erg  up to 1~Gpc 
(when the angular size of the cluster becomes comparable to the angular 
resolution of the  hard X-ray imager)  could be considered as potential targets for 
\astroh. 
Formally, the same is true also for $\gamma$-rays but,  because of the intergalactic absorption, the detectability  of  very distant  objects in TeV $\gamma$-rays is dramatically reduced.  
 
The expected $\gamma$-ray flux  from clusters of galaxies is at the limit of the sensitivity of present generation instruments. However it may be detectable with the future generation of detectors, in particular by CTA. The optimum energy interval for $\gamma$-ray detection  is between 1 and 10 TeV. Note that since the theoretical predictions  for the spectral shapes of synchrotron X-rays and IC $\gamma$-rays are  quite robust and depend rather weakly  on the model parameters, one can ``recover" the intrinsic spectrum of $\gamma$-rays  of energy $E \geq 10 \ \rm TeV$ based on the detected spectrum of hard X-rays. Then  the comparison of the recovered intrinsic  and detected  $\gamma$-ray spectra  should give us  an extremely important  information about the flux of the Extragalactic Background Light (EBL) at poorly know wavelengths between 10 and 100 $\mu$m. 

Finally we should comment on $\gamma$-rays related to interactions of accelerated protons with the ambient gas which can compete with the IC radiation of pair produced electrons. The relative contributions of these two channels depend on the density of the ambient gas and the spectral shape of accelerated protons.
The flux of $\gamma$-rays from {\it pp} interactions can be easily  estimated based on the cooling time of protons, $t_{\rm pp} \approx 1.5 \times 10^{19}/n_{-4}$ s, where $n_{-4}=n/10^{-4}$ cm$^{-3}$ is the density of the ambient hydrogen gas, normalised to $10^{-4}$ cm$^{-3}$.  Then, the energy flux of $\gamma$-rays at 1~TeV is estimated as $F_\gamma~(\sim 1~{\rm TeV})~\approx6\times10^{-12}\kappa W_{62} n_{-4}/d_{100}^2$~erg~cm$^{-2}$~s$^{-1}$, where $\kappa$ is the fraction of the total energy of accelerated protons in the energy interval between 10 to 100~TeV  (these protons are primarily responsible for production of $\gamma$-rays of energy $\sim 1$~TeV). For a proton energy spectrum extending to $10^{18}$ eV, this fraction is of order of $\kappa \sim 0.1$. Thus for an average gas density in a cluster like Coma, $n \sim 3 \times 10^{-4}$ cm$^{-3}$, the $\gamma$-ray flux at 1~TeV is expected at the level of $10^{-12} \  \rm erg\ cm^{-2}\ s^{-1}$ which is comparable to the contribution of IC radiation of secondary electrons. In the case of harder spectra of protons accelerated by non-linear shocks, the contribution of $\gamma$-rays from {\it pp} interactions is dramatically reduced and the contribution of  
secondary pairs to $\gamma$-rays via IC scattering  strongly dominates over $\gamma$-rays from {\it pp} interactions. Correspondingly, the upper limits obtained at GeV and TeV energies by the \lat\  and the Cherenkov telescopes 
\citep{Fermi_Stacked_Clusters,HESS-Coma,MAGIC_Perseus}  do not restrict the parameter space  for radiation of secondary electrons from interactions of ultrahigh energy cosmic rays with MBR.

\clearpage
\begin{multicols}{2}
{\footnotesize

\input{references}

}
\end{multicols}

\end{document}

%% file: ASTROHWhitePaperStyle.tex
\usepackage{ascmac}
\usepackage{amsmath}
\usepackage{amssymb}
\usepackage{bm}
\usepackage[footnotesize,bf]{caption}
\usepackage{fullpage}
\usepackage{color}
\usepackage{float}
\usepackage{graphicx}
\usepackage[hidelinks]{hyperref} 
\usepackage[]{natbib}
\usepackage{subfigure}
\usepackage{txfonts}
\usepackage{threeparttable}
\usepackage{wrapfig}
\usepackage[]{multicol}
\usepackage{wrapfig}
\usepackage{authblk}

\usepackage{floatflt}

\title{\large Complete list of the ASTRO-H Science Working Group}
\date{\vspace{-0.5cm}}
\newcommand{\MakeWhitePaperTitle}{
	\begin{center}
		\begin{figure}
			\vspace{1cm}
			\begin{center}
				\includegraphics[width=0.2\hsize]{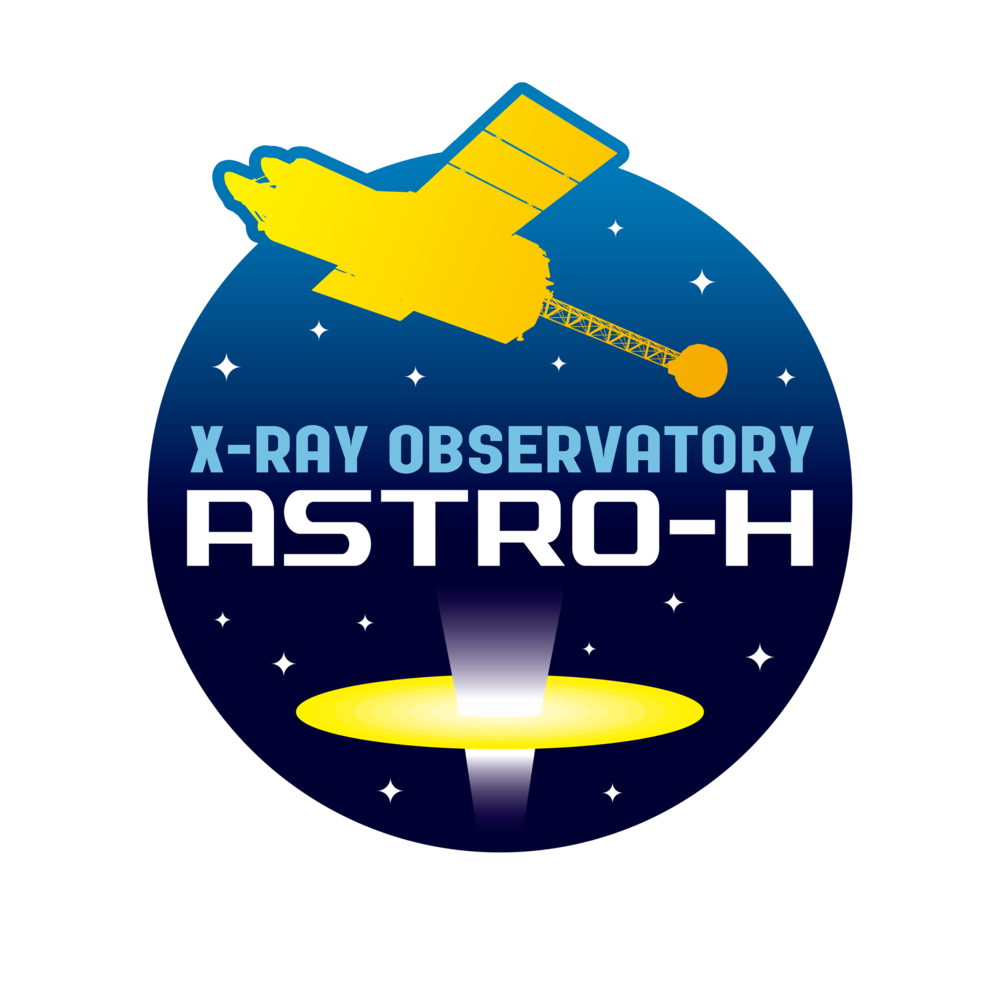}
			\end{center}
		\end{figure}
		\vspace{1cm}
		{\LARGE
		ASTRO-H Space X-ray Observatory\\
		White Paper\\
		}
		\vspace{5mm}
		{\large
		\WhitePaperTitle\\
		}
		\vspace{1cm}
		{
		\WhitePaperAuthors\\
		on behalf of the ASTRO-H Science Working Group
		}
	\end{center}
}

%% file: WP_AuthorList.tex
\usepackage{authblk}
\author[a]{Tadayuki~Takahashi}
\author[a]{Kazuhisa~Mitsuda}
\author[b]{Richard~Kelley}
\author[c]{Felix~Aharonian}
\author[d]{Hiroki~Akamatsu}
\author[e]{Fumie~Akimoto}
\author[f]{Steve~Allen}
\author[g]{Naohisa~Anabuki}
\author[b]{Lorella~Angelini}
\author[h]{Keith~Arnaud}
% ...
\affil[a]{Institute of Space and Astronautical Science (ISAS), Japan Aerospace Exploration Agency (JAXA), Kanagawa 252-5210, Japan}
\affil[b]{NASA/Goddard Space Flight Center, MD 20771, USA}
\affil[c]{Astronomy and Astrophysics Section, Dublin Institute for Advanced Studies, Dublin 2, Ireland}
\affil[d]{SRON Netherlands Institute for Space Research, Utrecht, The Netherlands}
\affil[e]{Department of Physics, Nagoya University, Aichi 338-8570, Japan}
\affil[f]{Kavli Institute for Particle Astrophysics and Cosmology, Stanford University, CA 94305, USA}
\affil[g]{Department of Earth and Space Science, Osaka University, Osaka 560-0043, Japan}
\affil[h]{Department of Astronomy, University of Maryland, MD 20742, USA}
% ...

%% file: myheader.tex

\def\s{\rule{0in}{0.23in}}
\def\ss{\rule{0in}{0.21in}}
\def\sun{\hbox{$\odot$}}
\def\la{\mathrel{\mathchoice {\vcenter{\offinterlineskip\halign{\hfil
$\displaystyle##$\hfil\cr<\cr\sim\cr}}}
{\vcenter{\offinterlineskip\halign{\hfil$\textstyle##$\hfil\cr
<\cr\sim\cr}}}
{\vcenter{\offinterlineskip\halign{\hfil$\scriptstyle##$\hfil\cr
<\cr\sim\cr}}}
{\vcenter{\offinterlineskip\halign{\hfil$\scriptscriptstyle##$\hfil\cr
<\cr\sim\cr}}}}}
\def\ga{\mathrel{\mathchoice {\vcenter{\offinterlineskip\halign{\hfil
$\displaystyle##$\hfil\cr>\cr\sim\cr}}}
{\vcenter{\offinterlineskip\halign{\hfil$\textstyle##$\hfil\cr
>\cr\sim\cr}}}
{\vcenter{\offinterlineskip\halign{\hfil$\scriptstyle##$\hfil\cr
>\cr\sim\cr}}}
{\vcenter{\offinterlineskip\halign{\hfil$\scriptscriptstyle##$\hfil\cr
>\cr\sim\cr}}}}}
\def\degr{\hbox{$^\circ$}}
\def\arcmin{\hbox{$^\prime$}}
\def\arcsec{\hbox{$^{\prime\prime}$}}
\def\utw{\smash{\rlap{\lower5pt\hbox{$\sim$}}}}
\def\udtw{\smash{\rlap{\lower6pt\hbox{$\approx$}}}}
\def\fd{\hbox{$.\!\!^{\rm d}$}}
\def\fh{\hbox{$.\!\!^{\rm h}$}}
\def\fm{\hbox{$.\!\!^{\rm m}$}}
\def\fs{\hbox{$.\!\!^{\rm s}$}}
\def\hour{\hbox{$^{\rm h}$}}
\def\min{\hbox{$^{\rm m}$}}
\def\fdg{\hbox{$.\!\!^\circ$}}
\def\dg{\hbox{$^{\rm d}$}}
\def\farcm{\hbox{$.\mkern-4mu^\prime$}}
\def\farcs{\hbox{$.\!\!^{\prime\prime}$}}

\def\aj{AJ}%
\def\araa{ARA\&A}%
\def\apj{ApJ}%
\def\apjl{ApJ}%
\def\apjs{ApJS}%
\def\ao{Appl.~Opt.}%
\def\apss{Ap\&SS}%
\def\aap{A\&A}%
\def\aapr{A\&A~Rev.}%
\def\aaps{A\&AS}%
\def\azh{AZh}%
\def\baas{BAAS}%
\def\jrasc{JRASC}%
\def\memras{MmRAS}%
\def\mnras{MNRAS}%
\def\pra{Phys.~Rev.~A}%
\def\prb{Phys.~Rev.~B}%
\def\prc{Phys.~Rev.~C}%
\def\prd{Phys.~Rev.~D}%
\def\pre{Phys.~Rev.~E}%
\def\prl{Phys.~Rev.~Lett.}%
\def\pasp{PASP}%
\def\pasj{PASJ}%
\def\qjras{QJRAS}%
\def\skytel{S\&T}%
\def\solphys{Sol.~Phys.}%
\def\sovast{Soviet~Ast.}%
\def\ssr{Space~Sci.~Rev.}%
\def\zap{ZAp}%
\def\nat{Nature}%
\def\iaucirc{IAU~Circ.}%
\def\aplett{Astrophys.~Lett.}%
\def\apspr{Astrophys.~Space~Phys.~Res.}%
\def\bain{Bull.~Astron.~Inst.~Netherlands}%
\def\fcp{Fund.~Cosmic~Phys.}%
\def\gca{Geochim.~Cosmochim.~Acta}%
\def\grl{Geophys.~Res.~Lett.}%
\def\jcp{J.~Chem.~Phys.}%
\def\jgr{J.~Geophys.~Res.}%
\def\jqsrt{J.~Quant.~Spec.~Radiat.~Transf.}%
\def\memsai{Mem.~Soc.~Astron.~Italiana}%
\def\nphysa{Nucl.~Phys.~A}%
\def\physrep{Phys.~Rep.}%
\def\physscr{Phys.~Scr}%
\def\planss{Planet.~Space~Sci.}%
\def\procspie{Proc.~SPIE}%

\def\jcap{J.~Cosm.~Astropart.~Phys.}%

\hyphenation{brems-strahl-ung}
